\documentclass[letterpaper,11pt]{article}
\usepackage{epsfig}

\newenvironment{appendletterA}
 {
  \setcounter{section}{0}
  \setcounter{equation}{0}
  
 }{
 }

\def\be{\begin{equation}}
\def\ee{\end{equation}}
\def\bc{\begin{center}}
\def\ec{\end{center}}
\def\bea{\begin{eqnarray}}
\def\eea{\end{eqnarray}}

\def\nn{\nonumber}
\def\ov{\overline}

\def\at{\alpha_t}
\def\ab{\alpha_b}
\def\as{\alpha_s}
\def\atau{\alpha_{\tau}}
\def\oat{{\cal O}(\at)}
\def\oab{{\cal O}(\ab)}
\def\oatab{{\cal O}(\at\ab)}
\def\oatas{{\cal O}(\at\as)}
\def\oabas{{\cal O}(\ab\as)}
\def\oatababq{{\cal O}(\at\ab + \ab^2)}
\def\oatqatababq{{\cal O}(\at^2 + \at\ab + \ab^2)}
\def\oatasatq{{\cal O}(\at\as + \at^2)}

\def\oatq{{\cal O}(\at^2)}
\def\oabq{{\cal O}(\ab^2)}
\def\oatauq{{\cal O}(\atau^2)}
\def\oabatau{{\cal O}(\ab \atau)}
\def\oas{{\cal O}(\as)}

\def\mstu{m_{\tilde{t}_1}^2}
\def\mstd{m_{\tilde{t}_2}^2}

\def\cdt{c_{2\bar{\theta}_t}}

\def\msbu{m_{\tilde{b}_1}^2}
\def\msbd{m_{\tilde{b}_2}^2}
\def\mgl{m_{\tilde{g}}}

\def\cdb{c_{2\bar{\theta}_b}}

\def\cptmptt{c_{\varphi_t-\tilde{\varphi}_t}}
\def\cpbmptb{c_{\varphi_b-\tilde{\varphi}_b}}

\def\cptpb{c_{\varphi_t+\varphi_b}}
\def\cpttptb{c_{\tilde{\varphi}_t+\tilde{\varphi}_b}}
\def\cptptb{c_{\varphi_t+\tilde{\varphi}_b}}
\def\cpbptt{c_{\varphi_b+\tilde{\varphi}_t}}

\def\mgut{M_{\scriptscriptstyle GUT}}
\def\m0{m_0}

\def\Veff{V_{\rm eff}}
\def\sigu{\Sigma_1}
\def\sigd{\Sigma_2}
\def\drbar{\overline{\rm DR}}

\def\msqu{m_{\tilde{q}_1}^2}
\def\msqd{m_{\tilde{q}_2}^2}
\def\msqi{m_{\tilde{q}_i}^2}
\def\thsq{\bar{\theta}_{\tilde{q}}}
\def\thq{\theta_{\tilde{q}}}
\def\Sdq{s_{2\theta_q}}

\def\tb{\tan\beta}
\def\mt{m_t}
\def\mb{m_b}
\def\mh{m_h}
\def\mH{m_{\scriptscriptstyle H}}
\def\mz{m_{\scriptscriptstyle Z}}

\def\ma{m_{\scriptscriptstyle A}}
\def\mabar{\ov{m}_{\scriptscriptstyle A}}
\def\mahat{\widehat{m}_{\scriptscriptstyle A}}
\def\mgut{M_{\scriptscriptstyle GUT}}

\def\sq2{\sqrt{2}}

\def\cb{c_{\beta}}
\def\sb{s_{\beta}}

\def\hlf{\frac{1}{2}}

\def\difft{\mstu-\mstd}
\def\diffb{\msbu-\msbd}
\def\C2b{c_{2\beta}}
\def\Sdt{s_{2\theta_t}}
\def\Sdb{s_{2\theta_b}}
\def\Cdt{c_{2\theta_t}}
\def\Cdb{c_{2\theta_b}}

\def\DVtu{\frac{\partial \Delta V}{\partial \mstu}}
\def\DVtd{\frac{\partial \Delta V}{\partial \mstd}}
\def\DVcdtq{\frac{\partial \Delta V}{\partial \cdt^2}}
\def\DVcptmptt{\frac{\partial \Delta V}{\partial \cptmptt}}
\def\DVcpbmptb{\frac{\partial \Delta V}{\partial \cpbmptb}}
\def\DVtt{\frac{\partial^{\,2} \Delta V}{(\partial \mt^2)^2 }}
\def\DVttu{\frac{\partial^{\,2} \Delta V}{\partial \mt^2 \partial \mstu }}
\def\DVttd{\frac{\partial^{\,2} \Delta V}{\partial \mt^2 \partial \mstd }}
\def\DVtutu{\frac{\partial^{\,2} \Delta V}{(\partial \mstu)^2 }}
\def\DVtutd{\frac{\partial^{\,2} \Delta V}{\partial \mstu \partial \mstd }}
\def\DVtdtd{\frac{\partial^{\,2} \Delta V}{(\partial \mstd)^2}}
\def\DVcdtqtu{\frac{\partial^{\,2} \Delta V}{\partial \cdt^2 \partial \mstu }}
\def\DVcdtqtd{\frac{\partial^{\,2} \Delta V}{\partial \cdt^2 \partial \mstd }}
\def\DVcdtqt{\frac{\partial^{\,2} \Delta V}{\partial \cdt^2 \partial \mt^2 }}
\def\DVcdtqcdtq{\frac{\partial^{\,2} \Delta V}{(\partial \cdt^2)^2 }}
\def\DVtub{\frac{\partial^{\,2} \Delta V}{\partial \mstu \partial \mb^2 }}
\def\DVtdb{\frac{\partial^{\,2} \Delta V}{\partial \mstd \partial \mb^2 }}
\def\DVbut{\frac{\partial^{\,2} \Delta V}{\partial \msbu \partial \mt^2 }}
\def\DVbdt{\frac{\partial^{\,2} \Delta V}{\partial \msbd \partial \mt^2 }}
\def\DVtb{\frac{\partial^{\,2} \Delta V}{\partial \mt^2 \partial \mb^2 }}
\def\DVtubu{\frac{\partial^{\,2} \Delta V}{\partial \mstu \partial \msbu }}
\def\DVtdbu{\frac{\partial^{\,2} \Delta V}{\partial \mstd \partial \msbu }}
\def\DVtubd{\frac{\partial^{\,2} \Delta V}{\partial \mstu \partial \msbd }}
\def\DVtdbd{\frac{\partial^{\,2} \Delta V}{\partial \mstd \partial \msbd }}
\def\DVtuc2b{\frac{\partial^{\,2} \Delta V}{\partial \mstu \partial \cdb^2 }}
\def\DVtdc2b{\frac{\partial^{\,2} \Delta V}{\partial \mstd \partial \cdb^2 }}
\def\DVtc2b{\frac{\partial^{\,2} \Delta V}{\partial \mt^2 \partial \cdb^2 }}
\def\DVbuc2t{\frac{\partial^{\,2} \Delta V}{\partial \msbu \partial \cdt^2 }}
\def\DVbdc2t{\frac{\partial^{\,2} \Delta V}{\partial \msbd \partial \cdt^2 }}
\def\DVbc2t{\frac{\partial^{\,2} \Delta V}{\partial \mb^2 \partial \cdt^2 }}
\def\DVc2tc2b{\frac{\partial^{\,2} \Delta V}{\partial \cdt^2 \partial \cdb^2}}
\def\DVcptpb{\frac{\partial \Delta V}{\partial \cptpb}}
\def\DVcpttptb{\frac{\partial \Delta V}{\partial \cpttptb}}
\def\DVcptptb{\frac{\partial \Delta V}{\partial \cptptb}}
\def\DVcpbptt{\frac{\partial \Delta V}{\partial \cpbptt}}

\begin{document}

\thispagestyle{empty}

\bc
\hfill{MPI-PhT/2003-21} \\
\hfill{TUM-HEP-507/03} \\
\hfill{RM3-TH/03-05}
\ec

\vspace{1.7cm}
\bc
{\LARGE\bf On the two--loop Yukawa corrections to the}
\ec

\bc
{\LARGE\bf MSSM Higgs boson masses at large $\tan\beta$} 
\ec

\vspace{1.4cm}

\bc

{\Large \sc A.~Dedes~$^{a,}$\footnote{\tt dedes@ph.tum.de},
G.~Degrassi~$^{b,}$\footnote{\tt degrassi@fis.uniroma3.it},
P.~Slavich~$^{c,\,d,}$\footnote{\tt slavich@mppmu.mpg.de}}\\
\vspace{1.2cm}

${}^a$
{\em Physik Department, Technische Universit\"at M\"unchen,\\
D--85748 Garching, Germany}
\vspace{.3cm}

${}^b$
{\em Dipartimento di Fisica, Universit\`a di Roma Tre and\\
INFN, Sezione di Roma III, Via della Vasca Navale 84, I--00146 Rome, Italy}
\vspace{.3cm}

${}^c$
{\em Institut f\"ur Theoretische Physik, Universit\"at Karlsruhe,\\
Kaiserstrasse 12, Physikhochhaus, D--76128 Karlsruhe, Germany}
\vspace{.3cm}

${}^d$
{\em Max Planck Institut f\"ur Physik,\\
F\"ohringer Ring 6, D--80805 M\"unchen, Germany}

\ec

\vspace{0.8cm}

\centerline{\bf Abstract}
\vspace{2 mm}
\begin{quote} \small
We complete the effective potential calculation of the two--loop,
top/bottom Yukawa corrections to the Higgs boson masses in the Minimal
Supersymmetric Standard Model, by computing the $\oatqatababq$
contributions for arbitrary values of the bottom Yukawa coupling.  We
also compute the corrections to the minimization conditions of the
effective potential at the same perturbative order. Our results extend
the existing $\oatq$ calculation, and are relevant in regions of the
parameter space corresponding to $\tan\beta \gg 1$. We extend to the
Yukawa corrections a convenient renormalization scheme, previously
proposed for the $\oabas$ corrections, that avoids unphysically large
threshold effects associated with the bottom mass and absorbs the bulk
of the corrections into the one--loop expression. For large values of
$\tan\beta$, the new contributions can account for a variation of
several GeV in the lightest Higgs boson mass.

\end{quote}
\vfill
\newpage
\setcounter{equation}{0}
\setcounter{footnote}{0}
\vskip2truecm


\section{Introduction}

One of main features of the Minimal Supersymmetric Standard Model
(MSSM)~\cite{revs} is the prediction of the existence of at least one
light Higgs boson~\cite{Haber}. After the conclusion of the LEP and
Tevatron Run I experimental programs that reported no significant
evidence for a Higgs boson, the experimental search for this particle
has now become one of the major tasks of the Tevatron Run II and of
the future LHC. Within the MSSM, the tree--level masses of the neutral
Higgs bosons can be parameterized in terms of three input parameters:
the mass of the CP--odd Higgs $\ma$, the Z boson mass $\mz$ and the
ratio of the two Higgs vacuum expectation values, $\tan\beta \equiv
v_2/v_1$. At tree level, at least one of the MSSM Higgs bosons is
bound to be lighter than the Z boson, thus the failure of detecting it
at LEP indicates that the MSSM could be a realistic theory only after
the radiative corrections to the Higgs boson masses have been taken
into account.

The radiative corrections arise from loop diagrams involving Standard
Model particles and their superpartners. Although the earliest
computations \cite{earlier} of radiative corrections to the MSSM Higgs
masses date back to the eighties, it was first realized in
Ref.~\cite{at} that the inclusion of the one--loop top/stop $\oat$
corrections, where $\alpha_t=h_t^2/(4\pi)$, $h_t$ being the
superpotential top coupling, may push the light Higgs mass well above
the tree--level bound. In the subsequent years, an impressive
theoretical effort has been devoted to the precise determination of
the MSSM Higgs masses: full one--loop computations have been
provided \cite{brignole,1loop}, leading logarithmic effects at two
loops have been included via appropriate renormalization group
equations \cite{rge,Apostolos}, and genuine two--loop corrections of
$\oatas$ \cite{hemphoang,Sven,Zhang,ezhat,dsz},
$\oatq$ \cite{hemphoang,ezhat,bdsz}, and $\oabas$ \cite{bdsz2} have
been evaluated in the limit of zero external momentum. The tadpole
corrections needed to minimize the effective potential, $\Veff$, have
also been calculated \cite{dstad} at the same perturbative
orders. Furthermore, the full two--loop corrections to the MSSM
effective potential have been calculated in Ref.~\cite{Martin1},
together with a first study of the effect of the two--loop corrections
to the Higgs masses controlled by the electroweak gauge couplings
\cite{Martin2}.

The corrections controlled by the top Yukawa coupling are in general
the most relevant ones. However, in regions of the MSSM parameter
space where $\tan\beta \gg 1$ the superpotential bottom coupling $h_b$
may be large (we recall that, at the classical level, $h_b / h_t =
(m_b / m_t) \tan\beta\,$) and the one--loop bottom/sbottom corrections
of $\oab$, where $\alpha_b=h_b^2/(4\pi)$, can be numerically relevant
and compete with those of $\oat$. At the two--loop level, the
evaluation of the corrections controlled by the bottom Yukawa coupling
requires the inclusion of one--loop, $\tan\beta$--enhanced threshold
corrections to the bottom mass \cite{hrs}. If the physical bottom mass
is used as input parameter in the one--loop part of the computation,
potentially large $\tan\beta$--enhanced corrections appear at two
loops. To address this problem, a set of renormalization prescriptions
for the parameters in the bottom/sbottom sector that avoid the
occurrence of unphysically large threshold effects at two loops was
proposed in Ref.~\cite{bdsz2} for the $\oabas$ part of the
corrections.

The purpose of this article is to complete the calculation of the
two--loop, top/bottom Yukawa corrections to the Higgs boson masses in
the effective potential approach. Such corrections were previously
computed \cite{hemphoang,ezhat,bdsz} in the limit \mbox{$h_b
\rightarrow 0$}, which is accurate enough only for moderate values of
$\tan\beta$. In that limit, the two--loop Yukawa corrections to the
MSSM Higgs masses are of ${\cal O}(\at^2 \mt^2)$, which we denote as
$\oatq$ for brevity. On the other hand, when the bottom Yukawa
coupling is left arbitrary, the resulting two--loop corrections are
proportional to various combinations of couplings and masses: e.g., we
find terms of ${\cal O}(\ab^2\,\mt^2)$, which might as well be
interpreted as $\tan\beta$--enhanced terms of ${\cal O} (\at
\ab\,\mb^2)$. To simplify our notation, we will refer to all such
``mixed'' terms as to $\oatab$ corrections, and to the terms that
depend only on the bottom Yukawa coupling as to $\oabq$
corrections. Our computation will thus provide us with the
$\oatqatababq$ corrections to the MSSM Higgs masses, extending the
$\oatq$ results obtained in Ref.~\cite{bdsz}. As a byproduct, we also
calculate the $\oatqatababq$ corrections to the minimization
conditions of the effective potential. We express our results in the
$\drbar$ renormalization scheme, as well as in an ``on--shell'' scheme
which extends the prescription described in Ref.~\cite{bdsz2} to the
case of the Yukawa corrections.  The resulting analytical formulae are
rather lengthy, thus we make them available, upon
request~\footnote{\tt E-mail: slavich@mppmu.mpg.de}, in the form of a
Fortran code.

The structure of this paper is the following. In section 2 we recall
some general issues of the effective potential approach to the
calculation of the Higgs masses. Section 3 describes our two--loop
computation of the $\drbar$ tadpoles and CP--odd, CP--even Higgs mass
matrices, while section 4 addresses our on--shell renormalization
prescription. Numerical results are given in section 5 and in section
6 we present a short discussion of the corrections controlled by the
tau Yukawa coupling as well as our conclusions.


\section{Higgs masses in the effective potential approach}

We begin our discussion by recalling some general results concerning the 
computation of the MSSM Higgs masses in the effective potential
approach.  The effective potential, which we write from the start in
terms of $\drbar$--renormalized fields and parameters, can be
decomposed as $\Veff = V_0 + \Delta V$, where $V_0$ is the tree--level
scalar potential and $\Delta V$ contains the radiative corrections.
Keeping only the dependence on the neutral Higgs fields $H_1^0$ and
$H_2^0$, the tree--level MSSM potential reads
\be
\label{V0}
V_0  =  (\mu^2+m_{H_1}^2) \, \left| H_1^0 \right|^2 
+ (\mu^2 +m_{H_2}^2) \, \left| H_2^0 \right|^2 
+ m_3^2 \, \left( H_1^0 H_2^0 + {\rm h.c.} \right)
+ {g^2 +g^{\prime\,2} \over 8} \left(
|H_1^0|^2 - |H_2^0|^2 \right)^2 \, ,
\ee
where: $\mu$ is the Higgs mass term in the superpotential (we assume it 
to be real, neglecting all possible CP--violating phases);
$m_{H_1}^2$, $m_{H_2}^2$ and $m_3^2$ are soft supersymmetry--breaking masses;
$g$ and $g'$ are the $SU(2)_L$ and $U(1)_Y$ gauge couplings,
respectively. The neutral Higgs fields can be decomposed into their
vacuum expectation values (VEVs) plus their CP--even and CP--odd
fluctuations as $H_i^0 = (v_i +S_i + i P_i)/\sq2 \;\;(i = 1,2)$. The
VEVs $v_i$ are determined by solving the minimization conditions of
the effective potential, i.e.
\be
\label{minimum}
\left.\frac{\partial\Veff}{\partial S_i} \right|_{\rm min}=0\,,
\;\;\;\;\;\;\;
\left.\frac{\partial\Veff}{\partial P_i} \right|_{\rm min}=0\,,
\ee
the second equality being automatically satisfied since we assume that
CP is conserved.  However, it is also possible to take $v_1$ and $v_2$
as input parameters, or equivalently $v^2 \equiv v_1^2 + v_2^2$ and
$\tan\beta \equiv v_2/v_1$, where $v^2$ is related to the squared
running mass of the $Z$ boson through $\mz^2 = (g^2 + g^{\prime
\,2})\,v^2/4$. In this case, the minimization conditions of $\Veff$
can be translated into conditions on $\mu^2$ and $m_3^2$:
\bea
\label{eqmu}
\mu^2 & = & - \frac{\mz^2}{2} 
+ \frac{m^2_{H_1} + \sigu - (m^2_{H_2} + \sigd) 
\,\tan^2\beta}{\tan^2\beta - 1}\,,\\
&\nn\\
\label{eqm3}
m_3^2 & = & \frac{\mz^2}{2}\,\sin 2\beta\, 
+ \hlf\,\tan 2\beta\, \left( m^2_{H_1} - m^2_{H_2} 
+ \sigu - \sigd \right) \, ,
\eea
where the ``tadpoles'' $\sigu$ and $\sigd$ are defined as
\be
\label{sigma}
\Sigma_i \equiv \frac{1}{v_i} \left.\frac{\partial \Delta V}{\partial S_i}
\right|_{\rm min}\,.
\ee
In the effective potential approach, the mass matrices for the neutral CP--odd
and CP--even Higgs bosons can be approximated by
\be
\label{defmat}
\left({\cal M}^2_P\right)^{\rm eff}_{ij} = 
\left. \frac{\partial^2  V_{\rm eff}}{\partial P_i \partial P_j}
\right|_{\rm min} \, , 
\hspace{1cm}
\left({\cal M}^2_S\right)^{\rm eff}_{ij} = 
\left. \frac{\partial^2  V_{\rm eff}}{\partial S_i \partial S_j}
\right|_{\rm min} \,.
\ee
Exploiting the minimization conditions of the effective potential,
Eq.~(\ref{minimum}), the CP--odd mass matrix can be written as
\be
\label{mpij}
\left({\cal M}^2_P\right)^{\rm eff}_{ij} = 
- m_3^2 \, \frac{v_1 v_2}{v_i v_j} - \delta_{ij}\,\Sigma_i 
+ \left.\frac{\partial^2 \Delta V}{\partial P_i \partial P_j}\right|_{\rm min}
\, .
\ee
$\left({\cal M}^2_P\right)^{\rm eff}$ has a
single non--vanishing eigenvalue that, in the approximation of zero
external momentum, can be identified with the squared physical mass of
the $A$ boson. We denote it as $\mabar^2 = \mahat^2 + \Delta \ma^2$,
where $\mahat^2 = -2\,m_3^2 /\sin 2\beta $ is the squared running mass  
of the $A$ boson. The CP--even mass matrix can in turn be decomposed 
as
\be
\left( {\cal M}^2_S \right)^{\rm eff}  = 
\left( {\cal M}^2_S \right)^{0, \, {\rm eff}} 
+
\left(\Delta{\cal M}^2_S\right)^{\rm eff}  \, ,
\label{effmatrix}
\ee
where the first term in the sum is the tree--level mass matrix
expressed in terms of $\mz$ and $\mabar$:
\be
\left( {\cal M}^2_S \right)^{0, \, {\rm eff}}
= 
\left(\begin{array}{cc}
\mz^2 \, \cb^2 + \mabar^2 \,\sb^2 
& -\left(\mz^2 + \mabar^2 \right) \sb\, \cb \\ 
-\left(\mz^2 + \mabar^2\right) \sb\, \cb 
& \mz^2 \, \sb^2 + \mabar^2 \,\cb^2 
\end{array}\right) 
\label{mpma} 
\, ,
\ee
($\cb \equiv \cos\beta\,,\;\sb \equiv \sin\beta$ and so on), while
the second term contains the radiative corrections:
\be
\label{dmsij}
\left(\Delta{\cal M}^2_S\right)_{ij}^{\rm eff}  = 
\left.\frac{\partial^2 \Delta V}{\partial S_i \partial S_j}
\right|_{\rm min} -(-1)^{i+j}\left.
\frac{\partial^2 \Delta V}{\partial P_i \partial P_j}\right|_{\rm min} \, .
\label{Dms}
\ee
It is clear from Eqs.~(\ref{mpij})--(\ref{dmsij}) that, in order to 
make contact with the physical $A$ mass, the effective potential should
be computed as a function of both CP--even and CP--odd fields.

Since $V_{\rm eff}$ generates one--particle--irreducible Green's
functions at vanishing external momentum, it is clear that the
effective potential approach neglects the momentum--dependent effects
in the Higgs self--energies. The complete computation of the physical
masses of the CP--even Higgs bosons, $\mh$ and $\mH$, and of the
CP--odd Higgs boson, $\ma$, requires the full, momentum--dependent
two--point functions. A detailed discussion of the correspondence
between the effective potential approach and the full computation has
been presented in Ref.~\cite{bdsz}. Here we just notice that the main
conclusions presented in that paper regarding the $\oatas$ and $\oatq$
corrections apply also to $\oatababq$ corrections discussed
here. Namely, the two--loop $\oatababq$ corrections to the lightest
Higgs eigenvalue are fully accounted for by the two--loop effective
potential evaluation of $m_h$ supplemented by known
momentum--dependent one--loop contributions, and the same is true for
$\mH$ when $\ma$ is not too large. Instead, if $\ma > m_t$ a full
two--loop $\oatababq$ computation of $\mH$ requires additional
momentum--dependent two--loop contributions, neglected by the
effective potential calculation, that have not been computed so far.


\section{Computation of the two--loop Yukawa corrections}

We shall now describe our two--loop computation of the tadpoles
$\Sigma_i\,$, the $A$--boson mass correction $\Delta \ma^2$ and the
matrix $\left(\Delta{\cal M}^2_S\right)^{\rm eff}$, including terms
controlled by the top and/or the bottom Yukawa couplings. The
computation is consistently performed in the gaugeless limit, i.e.~by
setting to zero all the gauge couplings, and by keeping $h_t$ and
$h_b$ as the only non--vanishing Yukawa couplings. In this limit, the
tree--level (field--dependent) spectrum of the MSSM simplifies
considerably: gauginos and Higgsinos do not mix; charged and neutral
Higgsinos combine into Dirac spinors with degenerate mass eigenvalues
$|\mu|^2$; the only massive SM fermions are the top and bottom quarks,
while all other fermions and gauge bosons have vanishing masses; the
only sfermions with non--vanishing couplings are the stop and sbottom
squarks; the lightest CP--even Higgs boson, $h$, is massless, and the
same is true for the Goldstone bosons; all the remaining Higgs states,
$(H,A, H^\pm)$, have degenerate mass eigenvalues $\ma^2$. The
tree--level mixing angle in the CP--even sector is just
$\alpha=\beta-\pi/2$.

The renormalization of the effective potential is performed according
to the lines of Ref.~\cite{dstad}, i.~e.~we express $\Veff$, from the
beginning, in terms of $\drbar$--renormalized fields and
parameters. In practice, this amounts to dropping all the divergent
terms in $\Delta V$ and replacing the two--loop integrals
$I(m_1^2,m_2^2,m_3^2)$ and $J(m_1^2,m_2^2)$ (see
e.~g.~Ref.~\cite{dstad} for the definitions) with their ``subtracted''
counterparts $\hat{I}$ and $\hat{J}$, first introduced in
Ref.~\cite{jackjones}.  Alternatively, we could follow the procedure
of Refs.~\cite{dsz,bdsz}: express $\Delta V$ in terms of bare
parameters and then renormalize the derivatives of $\Delta V$
(i.~e.~the tadpoles and the corrections to the Higgs masses), checking
explicitly the cancellation of the divergent terms. The general
formulae for the tadpoles and the corrections to the Higgs masses
would look slightly more complicated in the latter case. However, we
have checked that the two renormalization procedures lead to the same
final result, as they should.

According to Eqs.~(\ref{sigma}), (\ref{mpij}) and (\ref{dmsij}), the
tadpoles and the corrections to the Higgs mass matrices can be
computed by taking the derivatives of $\Delta V$ with respect to the
CP--even and CP--odd fields, evaluated at the minimum of
$\Veff$. Following the strategy of Refs.~\cite{dsz,bdsz}, we compute
$\Delta V$ in terms of a set of field--dependent parameters (masses
and angles), and use the chain rule to express the corrections in
terms of derivatives of $\Delta V$ with respect to those parameters.
In each sector, the field--dependent parameters can be chosen as
\be
\label{listpar}
m_q\,,\;\;\msqu\,,\;\;\msqd\,,\;\;\thsq\,,\;\;
\varphi_q\,,\;\;\widetilde{\varphi}_q \,,
\hspace{1cm} (q = t,b)\;,
\ee
where: $m_q$ and $\msqi$ are the quark and squark masses; $\thsq$ is
the field--dependent squark mixing angle, defined in such a way that
$0\leq \thsq < \pi/2$ (to be contrasted with the usual
field--independent mixing angle $\thq$, such that $-\pi/2 \leq \thq <
\pi/2$); $\varphi_q$ is the phase in the complex quark mass;
$\widetilde{\varphi}_q$ is the phase in the off--diagonal element of
the squark mass matrix. For the explicit Higgs field dependence of
these parameters, see Refs.~\cite{dsz,bdsz}. In the expression of
$\Delta V$ relevant to the $\oatq$ corrections (i.~e., with $h_b$ set
to zero), the top and stop phases always combine in the difference
$\varphi_t - \widetilde{\varphi}_t$, so that a convenient choice for
the field--dependent parameter is $\cptmptt \equiv \cos(\varphi_t -
\widetilde{\varphi}_t)$. On the other hand, when both $h_t$ and $h_b$
are nonzero, as it is the case in the present computation of the
$\oatqatababq$ corrections, the situation becomes more complicated:
besides the terms involving $\varphi_t - \widetilde{\varphi}_t$ and
$\varphi_b - \widetilde{\varphi}_b$, we find other terms, coming from
diagrams with a charged Higgs or Goldstone boson, that involve the
combinations $\varphi_t + \widetilde{\varphi}_b\,,\; \varphi_b +
\widetilde{\varphi}_t\,,\;\varphi_t +\varphi_b$ and
$\widetilde{\varphi}_t+\widetilde{\varphi}_b$.

Exploiting the field--dependence of the various masses and angles, we
get the following general formulae for the $\oatqatababq$ corrections
in the $\drbar$ renormalization scheme:

\bea
\label{dms11}
\left(\Delta {\cal M}^2_S\right)^{\rm eff}_{11} & = & 
2 \, h_b^2\, m_b^2\, F^b_1
+ 2\, h_b^2\, A_b\, m_b\, \Sdb\, F^b_2 
+ \hlf \, h_b^2\, A_b^2\, \Sdb^2\, F^b_3 \nn\\
&+& 
\hlf \, h_t^2 \,\mu^2 \,\Sdt^2  \,F^t_3
+ 2 \,h_t\,h_b\, \mb \, \mu \,  \Sdt \, F^t_4
+ h_t\,h_b\, \mu \, A_b \,\Sdt\,\Sdb\, F_5\,, \\
&&\nn\\
\label{dms12}
\left(\Delta {\cal M}^2_S\right)^{\rm eff}_{12} & = &
h_t^2 \,\mu\, m_t\, \Sdt \,  F^t_2 
+ \hlf \, h_t^2\, A_t \,\mu \, \Sdt^2 \, F^t_3
+ h_t\,h_b\, \mb \, A_t \, \Sdt \, F^t_4\nn\\
&+& h_b^2 \,\mu\, m_b\, \Sdb \,  F^b_2 
+ \hlf \, h_b^2\, A_b \,\mu \, \Sdb^2 \, F^b_3 
+ h_t\,h_b\, \mt \, A_b \, \Sdb \, F^b_4\nn\\
&+& \hlf \,h_t\,h_b\,\Sdt \,\Sdb \,(A_t\,A_b + \mu^2)\,F_5
+ 2 \,h_t\,h_b\, \mt \, \mb \, F_6\,, \\
&&\nn\\
\label{dms22}
\left(\Delta {\cal M}^2_S\right)^{\rm eff}_{22} & = &
2\, h_t^2\, m_t^2\, F^t_1
+ 2\, h_t^2\, A_t\, m_t\, \Sdt\, F^t_2 
+ \hlf \, h_t^2\, A_t^2\, \Sdt^2\, F^t_3\nn\\
&+& \hlf \, h_b^2 \,\mu^2 \,\Sdb^2  \,F^b_3
+ 2 \,h_t\,h_b\, \mt \, \mu \,  \Sdb \, F^b_4
+ h_t\,h_b\, \mu \, A_t \,\Sdt\,\Sdb\, F_5\,,\\
&&\nn\\
\label{sigma1}
v_1^2\,\Sigma_1 
& = & m_t\, \mu\, \cot\beta\, \Sdt\, F^t
+ \mb\, A_b\, \Sdb\, F^b + 2\, \mb^2\, G^b \,,\\
&&\nn\\
\label{sigma2}
v_2^2\,\Sigma_2
& = & \mb\, \mu\, \tan\beta\, \Sdb\, F^b
+m_t\, A_t\, \Sdt\, F^t + 2\, m_t^2\, G^t  \,,\\
&&\nn\\
\label{deltama}
\Delta m_A^2 & = &
-\frac{1}{\cb\,\sb}\,\left(\, 
\frac{h_t^2 \,\mu\,A_t}{\difft}\, F^t +
\frac{h_b^2 \,\mu\,A_b}{\diffb}\, F^b 
+ 2\,h_t\,h_b\,F_A\,\right)\;.
\eea
In the equations above, $A_t$ and $A_b$ are the soft supersymmetry--breaking
trilinear couplings of the Higgs fields to the stop and sbottom
squarks, and $\Sdq \equiv \sin 2\thq\;$ ($q=t,b$) refer to the usual
field--independent squark mixing angles. The functions $F_i^{q}\; (i =
1,2,3,4)\,,\; F_5\,,\; F_6\,,\; F^q\,,\; G^q$ and $F_A$ are
combinations of the derivatives of $\Delta V$ with respect to the
field--dependent parameters, computed at the minimum of the effective
potential; their definitions are given in the appendix. It can be
noticed that, as it is predictable from the form of the MSSM Lagrangian,
the above results are fully symmetric with respect to the simultaneous
replacements $t \leftrightarrow b$ and $H_1 \leftrightarrow H_2$, the
latter resulting into $\tan\beta \leftrightarrow \cot\beta\,,\; v_1
\leftrightarrow v_2\,,\; \left(\Delta{\cal M}^2_S\right)^{\rm
eff}_{11} \leftrightarrow \left(\Delta{\cal M}^2_S\right)^{\rm
eff}_{22}$ and $\sigu \leftrightarrow \sigd$.

An explicit expression of the two--loop top and bottom Yukawa
contribution to $\Delta V$ can be found in Ref.~\cite{ezhat}, while
the complete two--loop effective potential for the MSSM was given in
the second paper of Ref.~\cite{Martin1}. However, those expressions
were computed for vanishing CP--odd fields, thus omitting the
dependence on the phases $\varphi_q$ and
$\widetilde{\varphi}_q$. Since these phases appear in $\Delta V$ in
many different combinations, it is not possible to infer the general
field--dependent expression of $\Delta V$ by means of simple
substitutions in Eq.~(D.6) of Ref.~\cite{ezhat}, as it was the case in
the computation of the $\oatasatq$ corrections~\footnote{Also, we do
not agree with Ref.~\cite{ezhat} on the sign of the penultimate line
of Eq.~(D.6).}.

We worked out the general expression of the two--loop top and bottom
Yukawa contribution to $\Delta V$ in terms of all the field--dependent
parameters of Eq.~(\ref{listpar}). We then computed its derivatives in
order to obtain explicit formulae for the various functions appearing
in Eqs.~(\ref{dms11})--(\ref{deltama}). The use of a recursive
relation for the derivatives of $I(m_1^2,m_2^2,m_3^2)$, presented in
Ref.~\cite{dstad}, helped us to keep the number of terms involved
under control. However, the resulting analytical formulae are very
long and we choose not to display them in print. Instead, we make
them available in the form of a Fortran code.


\section{On--shell renormalization scheme and input parameters}

The results presented in the previous section are valid when the MSSM
input parameters are expressed in the $\drbar$ renormalization
scheme. This way of presenting the results is convenient for analyzing
models that predict, via the MSSM renormalization group equations, the
low--energy $\drbar$ values of the parameters in terms of a set of
boundary conditions assigned at some scale $\mgut$ much larger than
the weak scale (see Refs.~\cite{rgecodes,ASBS} for a list of public
codes that are commonly used in this kind of analyses, and
Ref.~\cite{kraml} for a comparison among them). General low--energy
analyses of the MSSM, however, do not refer to boundary conditions at
high scales, and are usually performed in terms of parameters with a
more direct physical interpretation, such as pole masses and
appropriately defined mixing angles in the squark sector. Such an
approach requires modifications of our two--loop results, induced by
the variation of the parameters entering the one--loop corrections
when moving from the $\drbar$ scheme to a different scheme (for a
generic parameter $x$, we define the shift from the $\drbar$ value
$\hat{x}$ as $\delta x \equiv \hat{x} - x$).

However, it is not always possible to find a sensible definition with
a direct physical interpretation for all the relevant parameters.  For
example, while there is a well known physical ($\equiv$ pole)
definition for the masses, the so--called ``On--Shell'' (OS)
definition, and an OS definition for the squark mixing angles can be
also conceived \cite{guasch}, it is not clear what meaning should be
assigned to an OS definition of parameters like $(A_t\,,\,A_b\,,\,
\mu\,,\,\tan\beta)$. For instance, they could be related to specific
physical amplitudes. However, given our present ignorance of any
supersymmetric effect, such a choice does not seem particularly
useful. In these cases it seems sometimes simpler to stick to a
$\drbar$ definition.

It is rather easy to devise an OS renormalization scheme for the
parameters in the top/stop sector, based on the OS prescription for
the top and stop masses and the stop mixing angle and treating $A_t$
as a derived quantity, while retaining a $\drbar$ definition for $\mu$
and $\tan\beta$ (see e.~g.~Refs.~\cite{dsz,bdsz}).  Instead, some
additional care is required in the choice of an OS scheme for the
parameters in the bottom/sbottom sector, due to the potentially large
one--loop threshold corrections \cite{hrs}, proportional to
$\tan\beta$, that contribute to the pole bottom mass.  For example, a
definition of $A_b$ in terms of the OS bottom and sbottom masses and
sbottom mixing angle, similar to the definition of $A_t$, would
produce a shift $\delta A_b$ proportional to $\tan^2\beta$
\cite{eberl}. When $\tan\beta$ is large, this would induce very large
corrections to the Higgs masses at two loops, questioning the validity
of the perturbative expansion.

To overcome this problem, we adopt a set of renormalization
prescriptions for the parameters in the the bottom/sbottom sector,
first introduced in Ref.~\cite{bdsz2} for the case of the strong
corrections, that avoid the occurrence of unphysically large
threshold effects and at the same time enforce other desirable
properties such as the decoupling of heavy particles, the infrared
finiteness and gauge--independence. Generalizing these prescriptions to
the case of the Yukawa corrections, and combining them with the usual
prescriptions for the top/stop parameters \cite{bdsz}, we obtain a
convenient OS renormalization scheme for the $\oatababq$ part of the
corrections to the Higgs masses.  Since the corrections controlled by
the bottom Yukawa coupling can be sizeable only for large values of
$\tan\beta$, we work directly in the physically relevant limit of
$\tan\beta \rightarrow \infty$, i.~e.~$v_1 \rightarrow 0\,, \;v_2
\rightarrow v$.

For the OS squark masses and mixing angles, top quark mass and
electroweak parameter $v\equiv (\sq2\,G_{\mu})^{-1/2}$ we adopt the
definitions
\be
\label{ospars}
\delta m^2_{\tilde{q}_i} = \Pi_{ii}^{\tilde{q}}(m^2_{\tilde{q}_i})
\,,\;\;\;\;
\delta \theta_{\tilde{q}} = \hlf \,
\frac{\Pi_{12}^{\tilde{q}}(\msqu) + \Pi_{12}^{\tilde{q}}(\msqd)}{\msqu-\msqd}
\,,\;\;\;\;
\delta\mt = \Sigma_t(\mt)\,,\;\;\;\;
\delta v = \frac{v}{2}\,\frac{\Pi_{WW}^{T}(0)}{m_W^2}\,,
\ee
where $\tilde{q} = (\tilde{t}\,,\tilde{b})\,,$ while
$\Pi^{\tilde{q}}_{ij}(p^2)\,,\; \Sigma_t(p)$ and $\Pi_{WW}^{T}(p^2)$
denote the real and finite parts of the self--energies of squarks, top
quark and $W$ boson, respectively. Following Ref.~\cite{bdsz}, we
further treat $\mu$ as a $\drbar$ parameter computed at a reference
scale $Q_0 = 175$ GeV, and $h_t$ and $A_t$ as derived quantities that
can be computed by means of the tree--level formulae for $\mt$ and
$\Sdt$, respectively.
In principle, we still have to define $\mb\,,\, h_b$ and $A_b$.
However, in the large $\tan\beta$ limit, the bottom mass is just zero,
and the sbottom mixing angle becomes
\be
\label{lims2b}
\Sdb \; = \;\frac{\sqrt{2} \, h_b \, \mu \,
v }{m_{\tilde{b}_1}^2 - m_{\tilde{b}_2}^2} \,, 
\ee
which is independent of $\mb$ and $A_b\,$. We can thus treat $h_b$ as
a quantity derived from the sbottom mixing, and use Eqs.~(\ref{ospars})
and (\ref{lims2b}) to obtain a prescription for $\delta h_b$:
\be
\label{dhb}
\delta h_b = h_b\,\left(
\frac{\delta\msbu - \delta\msbd}{\msbu-\msbd}
+ \frac{\delta\Sdb}{\Sdb} - \frac{\delta v}{v}\right)\,.
\ee
In Ref.~\cite{bdsz2}, an OS definition for the quantity
$\widetilde{A}_b \equiv h_b\, A_b$, or equivalently for $\delta A_b =
(\delta\widetilde{A}_b - \delta h_b \, A_b)/h_b\,$, was proposed in terms of 
the $(\widetilde{b}_1 \widetilde{b}_2^* A)$ proper vertex $i
\Lambda_{12A}(p_1^2, p_2^2, p_A^2)$ for the case of the strong
corrections.  A generalization of that definition that can also
encompass the Yukawa corrections is given by
\bea
\label{deltaab}
\delta\widetilde{A}_b & = & - {i \over \sqrt{2}}\left[
\Lambda_{12A}(\msbu,\msbu,0) + \Lambda_{12A}(\msbd,\msbd,0) \right] \nn\\
&&\nn\\
&+ &  \frac{\widetilde{A}_b}{2} \, \left[\, 
\frac{\Pi_{11}^{\tilde{b}}(\msbu) - \Pi_{11}^{\tilde{b}}(\msbd)}{\msbu- \msbd}
+\frac{\Pi_{22}^{\tilde{b}}(\msbu) - \Pi_{22}^{\tilde{b}}(\msbd)}{\msbu- \msbd}
+\frac{\Pi_{AA}(\msbu) - \Pi_{AA}(\msbd)}{\msbu- \msbd}
\,\right]\,.\nn\\
\eea
Having fully specified our OS renormalization prescriptions in the
limit $\tan\beta \rightarrow \infty$, physically relevant for the
$\oatababq$ corrections, we can proceed to obtain formulae for the
CP--even Higgs masses in our OS scheme and merge them with the known $\oatq$ OS
results \cite{bdsz} that contain an explicit dependence on
$\tan\beta$. This can be done in three steps: first, we take the limit
of $\tan\beta \rightarrow \infty\,,\;\mb \rightarrow 0$ in the general
$\drbar$ results for the $\oatqatababq$ part of 
$\left(\Delta{\cal M}^2_S\right)^{\rm eff}$; then we add the
contributions due to the shifts of the parameters entering the
one--loop corrections (this requires the computation of the ${\cal
O}(\at+\ab)$ part of the counterterms in the large $\tan\beta$ limit);
finally, we subtract from this results the pure $\oatq$ part which,
being relevant for all values of $\tan\beta$, must instead be computed
separately with the formulae of Ref.~\cite{bdsz}. Notice that we do
not encounter any terms that blow up when we take the limit of large
$\tan\beta$ in the $\drbar$ results: unphysically large contributions
could only be introduced by hand, as the result of a poor choice of
the renormalization conditions for the parameters in the
bottom/sbottom sector.

We discuss now the parameters that we will actually use as inputs of
our calculation.  In particular, although we have used
Eqs.~(\ref{lims2b})--(\ref{dhb}) to define an OS bottom Yukawa
coupling $h_b$ through the sbottom mixing, we still need to exploit
the experimental information on the bottom mass in order to obtain the
$\drbar$ running coupling $\hat{h}_b$.  The OS coupling will then be
computed through the relation $h_b = \hat{h}_b - \delta
h_b$. Following Ref.~\cite{bdsz2}, we define the running coupling
$\hat{h}_b$ at the reference scale $Q_0 = 175$ GeV to be
\be
\hat{h}_b \equiv h_b(Q_0)_{\rm MSSM}^{\drbar} =  
{\ov{m}_b \sqrt{2} \over v_1} {1 + \delta_b \over \left| 1 +
\epsilon_b \right| } \, ,
\label{ourhb}
\ee
where: $ \ov{m}_b \equiv m_b(Q_0)_{\rm SM}^{\drbar} = 2.74 \pm 0.05$
GeV is the Standard Model bottom mass, evolved up to the scale $Q_0$
to take into account the resummation of the universal large QCD
logarithms; $\epsilon_b$ contains the $\tan\beta$--enhanced threshold
corrections from both the gluino--sbottom and the higgsino--stop loops
(denoted as $\epsilon_b$ and $\epsilon_b^{\,\prime}\,,$ respectively,
in Eqs.~(25) and (26) of Ref.~\cite{bdsz2}); $\delta_b$ contains the
residual threshold corrections that are not enhanced by
$\tan\beta$. Notice that, as shown in Ref.~\cite{resum}, keeping
$\epsilon_b$ in the denominator of Eq.~(\ref{ourhb}) allows to resum
the $\tan\beta$--enhanced threshold corrections to all orders in the
perturbative expansion. On the other hand, there is no preferred way
of including the threshold corrections parametrized by $\delta_b$,
whose effect on the value of $\hat{h}_b$ is anyway very small.
Neglecting all the terms controlled by the electroweak gauge 
couplings, $\epsilon_b$ reads
\bea
\epsilon_b  & = &  
-\frac{2 \,\alpha_s}{3 \, \pi} \,\frac{\mgl\,\mu\,\tan\beta}
{ \msbu-\msbd}\,\left[ 
\frac{\msbu}{\msbu - \mgl^2} \,\ln \frac{\msbu}{\mgl^2} -
\frac{\msbd}{\msbd - \mgl^2} \,\ln \frac{\msbd}{\mgl^2}
\right]\nn\\
&&\nn\\
& & 
-\frac{\at}{4\,\pi} \, \frac{A_t\,\mu\,\tan\beta}
{\mstu-\mstd}\,\left[ 
\frac{\mstu}{\mstu - \mu^2 }\,\ln \frac{\mstu }{ \mu^2} -
\frac{\mstd}{\mstd - \mu^2 }\,\ln \frac{\mstd }{ \mu^2}
\right]\,.
\label{epsilon}
\eea
It appears from Eq.~(\ref{ourhb}) that $\hat{h}_b$ blows up when
$\epsilon_b$ approaches $-1$, in which case the correct value of the
bottom mass cannot be reproduced with $\hat{h}_b$ in the perturbative
regime, and the corresponding set of MSSM parameters must be
discarded. It can also be noticed from Eq.~(\ref{epsilon}) that, since
we take $\mgl >0\,,$ for $A_t >0$ ($A_t <0$) the $\oas$ and $\oat$
contributions enter $\epsilon_b$ with the same (the opposite)
sign. Moreover, if we take only the $\oas$ part of $\epsilon_b$ into
account, $\hat{h}_b$ can be enhanced by the threshold correction only
for large values of $\mgl$ and large and positive~\footnote{Our
convention for the sign of $\mu$ is such that, e.~g., the sbottom
mixing parameter reads $X_b = A_b + \mu\,\tan\beta$.}  values of
$\mu$, whereas, when we include the $\oat$ part, $\hat{h}_b$ can be
enhanced also for small values of $\mgl$ and large and negative values
of both $\mu$ and $A_t$.

For the top/stop sector, we take as input the current central value of
the top pole mass, $m_t = 174.3$ GeV \cite{mtop}, and the parameters
$(m_{Q,\tilde{t}}\,, m_U, A_t)$ that can be derived by rotating the
diagonal matrix of the OS stop masses by the angle
$\theta_{\tilde{t}}$, defined in Eq.~(\ref{ospars}). Concerning the
sbottom sector, additional care is required, because of our
non--trivial definition of $h_b$ and of the fact that, at one loop,
the parameter $m_{Q,\tilde{b}}$ entering the sbottom mass matrix
differs from the corresponding stop parameter $m_{Q,\tilde{t}}$ by a
finite shift \cite{eberl}.  We start by computing the renormalized
coupling $h_b$ as given by Eq.~(\ref{dhb}) and (\ref{ourhb}). Then we
compute $m_{Q,\tilde{b}}$ following the prescription of
Ref.~\cite{eberl}.  Finally, we use the parameters $h_b$ and
$m_{Q,\tilde{b}}$ to compute the actual values of the OS sbottom
masses and mixing angle. Concerning the $A$--boson mass, which enters
the tree--level mass matrix for the CP--even Higgses, we take as input
the physical mass $\ma$, dropping the distinction between $\ma$ and
the effective potential mass $\mabar$ (this amounts to neglecting the
effect of the uncomputed momentum--dependent two--loop
corrections). The renormalization of the $Z$--boson mass, whose
numerical value we take equal to the physical mass $\mz = 91.187$ GeV,
does not affect the $\oatababq$ corrections.  The remaining numerical
inputs are the OS electroweak parameter $v = 246.218$ GeV and the
strong coupling constant, which we fix as $\as(Q_0) = 0.108$.


\section{Numerical results} 

We are now ready to discuss the numerical effect of our two--loop
corrections. In the previous sections we have discussed how to express
our results in either the $\drbar$ renormalization scheme or an OS
scheme suitably chosen to separate the genuine two--loop corrections
from the threshold corrections in the relation between $\mb$ and
$h_b$. Our $\drbar$ results for the two--loop corrections to the Higgs
masses and to the electroweak symmetry breaking conditions can be
easily implemented in the existing codes \cite{rgecodes,ASBS} that
compute the MSSM mass spectrum from a set of unified parameters at the
scale $\mgut$. A study of the implications of our results in the
framework of gravity (mSUGRA), gauge (GMSB) or anomaly (AMSB) mediated
supersymmetry breaking models goes beyond the scope of this paper, and
will appear elsewhere \cite{djouadi}. In the following discussion, we
will adopt a low--energy point of view and assume that the various
input parameters are related, when possible, to physical
quantities. To this aim, we will make use of the OS renormalization
scheme presented in section 4. We recall that, although our OS
prescription is defined in the limit $\tan\beta \rightarrow \infty$,
the corrections have an indirect dependence on $\tan\beta$ coming from
the input value for $\hat{h}_b\,,$ see Eq.~(\ref{ourhb}).

\begin{figure}[p]
\begin{center}
\epsfig{figure=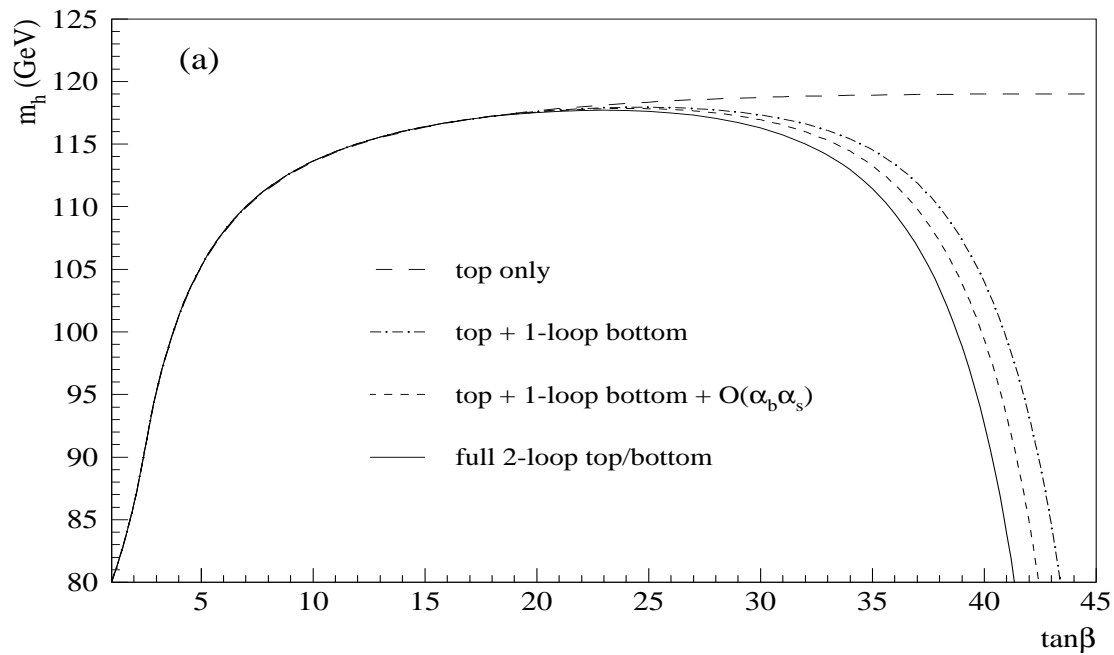,width=16cm,height=10cm}
\epsfig{figure=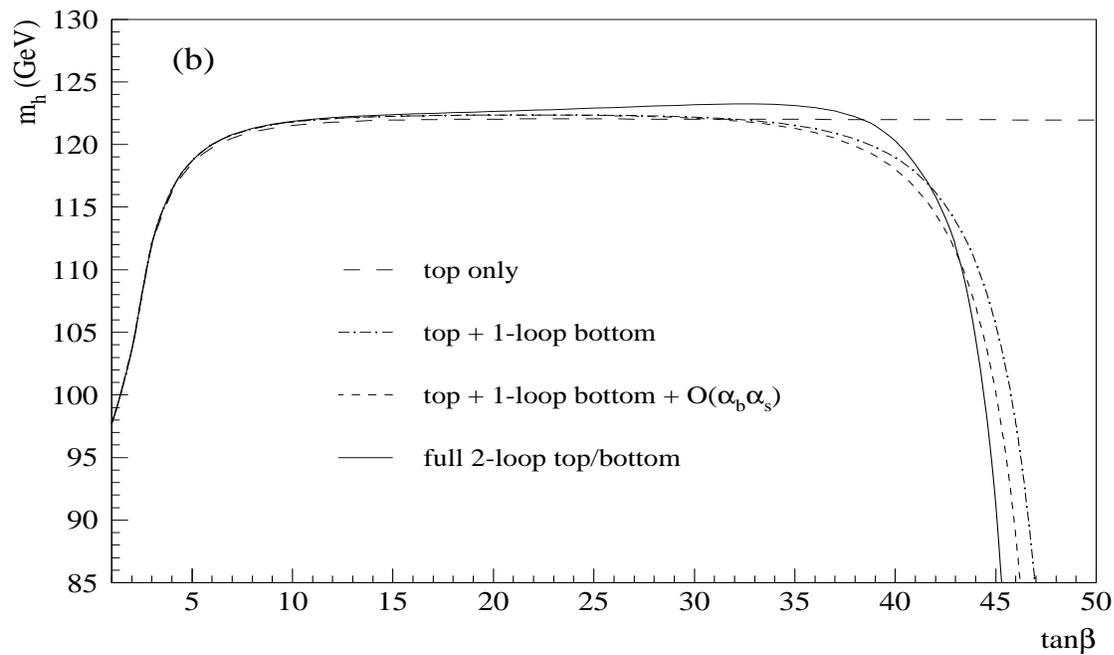,width=16cm,height=10cm}
\end{center}
\caption{The mass $m_h$ as a function of $\tan\beta$, for $\ma = 120$
GeV (upper panel) or 250 GeV (lower panel).  The other input
parameters are $A_t = 1$ TeV, $A_b = 2$ TeV, $\mu = m_{Q,\tilde{t}} =
m_U = m_D = \mgl = 1$ TeV.  The meaning of the different curves is
explained in the text.}
\label{figmhtb}
\end{figure}
In Figs.~\ref{figmhtb}a (upper panel) and \ref{figmhtb}b (lower panel)
we show the light Higgs mass $m_h$ as a function of $\tan\beta$, for
$\ma = 120$ GeV and $\ma = 250$ GeV, respectively. The other input
parameters are chosen as $A_t = 1$ TeV, $A_b = 2$ TeV, $\mu =
m_{Q,\tilde{t}} = m_U = m_D = \mgl = 1$ TeV. In each plot, the
long--dashed curve corresponds to the value of $m_h$ obtained at
${\cal O}(\at + \at \as + \at^2)$, i.~e.~by including only the one--
and two--loop corrections controlled by the top Yukawa coupling; the
dot--dashed curve includes in addition the one--loop $\oab$
corrections, controlled by the bottom Yukawa coupling~\footnote{In the
calculation of the one--loop $\oat$ and $\oab$ corrections we include the
effects proportional to $\mz^2$ and the momentum corrections as in
\cite{brignole}.}; the short--dashed curve includes the two--loop
$\oabas$ corrections computed in Ref.~\cite{bdsz2}; finally, the solid
curve corresponds to the full two--loop Yukawa computation of $m_h$,
i.~e.~it includes also the $\oatababq$ corrections discussed in the
previous sections. We can see from Figs.~\ref{figmhtb}a and
\ref{figmhtb}b that the corrections controlled by the top Yukawa
coupling depend very weakly on $\tan\beta$ when the latter is
large. On the other hand, the $\oab$ corrections lower considerably
$m_h$ when $\tan\beta$ increases. Concerning the two--loop corrections
controlled by the bottom Yukawa coupling, the comparison between the
dot--dashed and short--dashed curves shows that the $\oabas$
corrections amount to a small fraction of the $\oab$ ones, but they
can still lower $m_h$ by several GeV when $\tan\beta$ is large. The
comparison between the short--dashed and solid curves shows that the
effect of the $\oatababq$ corrections can also amount to several GeV
when $\tan\beta$ is large. From Fig.~\ref{figmhtb}a we see that, when
$\ma$ is small and the correction to $m_h$ is mainly driven by
$(\Delta {\cal M}_S^2)_{11}$, the $\oatababq$ corrections enter with
the same sign as the $\oabas$ corrections, reducing further the value
of $m_h$. On the other hand, Fig.~\ref{figmhtb}b shows that for larger
values of $\ma$, when the correction to $m_h$ is sensitive to $(\Delta
{\cal M}_S^2)_{22}$, the new corrections account for an increase in
$m_h$ of a few GeV at moderately large values of $\tan\beta$ (i.~e.,
$\tan\beta \approx$ 30--40). This is basically due to a positive
contribution to $(\Delta {\cal M}_S^2)_{22}$ coming from the $\oatab$
part of corrections. When $\tan\beta$ takes on larger values, however,
the overall effect of the $\oatababq$ corrections to $m_h$ turns again
to negative.

\begin{figure}[t]
\begin{center}
\epsfig{figure=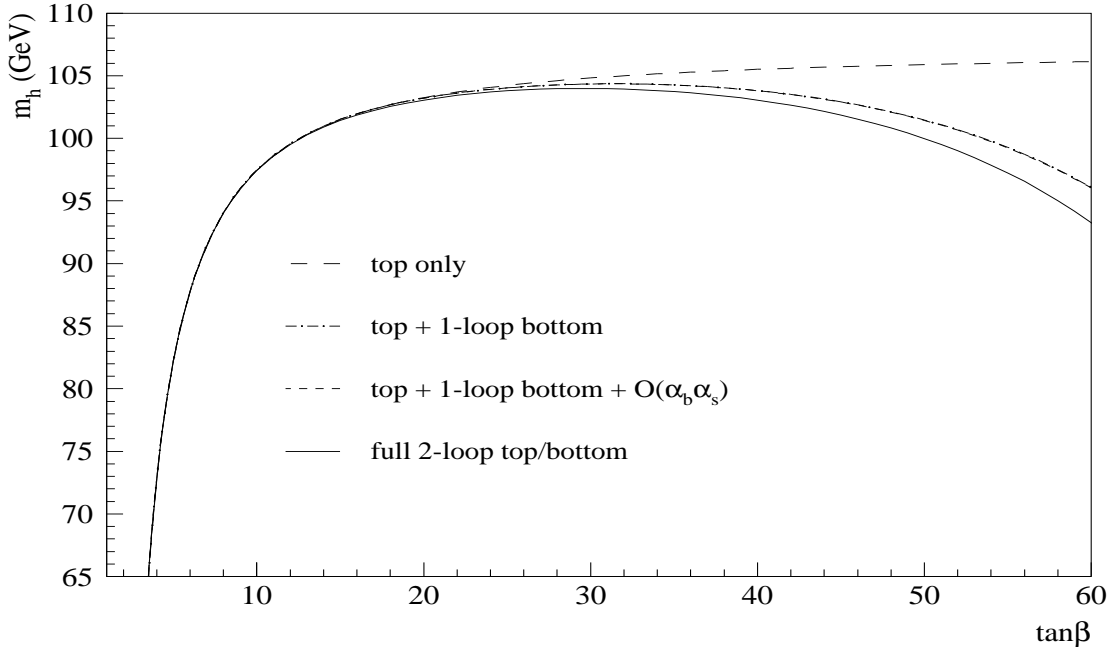,width=16cm,height=10cm}
\end{center}
\vspace{-0.6cm}
\caption{The mass $m_h$ as a function of $\tan\beta$, for $\ma = 120$
GeV, $\mu = A_t = -2$ TeV, $A_b = -3$ TeV, 
$m_{Q,\tilde{t}} = m_U = m_D = 1$ TeV
and $\mgl = 200$ GeV.  The meaning of the different curves is
explained in the text.}
\label{figmhtbglu}
\end{figure}

It is interesting to realize that the $\oatababq$ corrections can be
sizeable also for parameter choices that make the $\oabas$ corrections
irrelevant.  In Fig.~\ref{figmhtbglu} we show $m_h$ as a function of
$\tan\beta$, for $\ma = 120$ GeV, $\mu = A_t = - 2$ TeV, $A_b = - 3$ TeV,
$m_{Q,\tilde{t}} = m_U = m_D = 1$ TeV and $\mgl = 200$ GeV. The
meaning of the various curves is the same as in Fig.~\ref{figmhtb}.
Due to the small value of the gluino mass with respect to the sbottom
masses, the $\oabas$ corrections to $m_h$ are negligible (in fact, the
dot--dashed and short--dashed curves overlap). On the other hand,
comparing the short--dashed and solid curves we see that the
$\oatababq$ corrections can still amount to a few GeV when $\tan\beta$
is large enough.

\begin{figure}[t]
\begin{center}
\vspace{-0.5cm}
\epsfig{figure=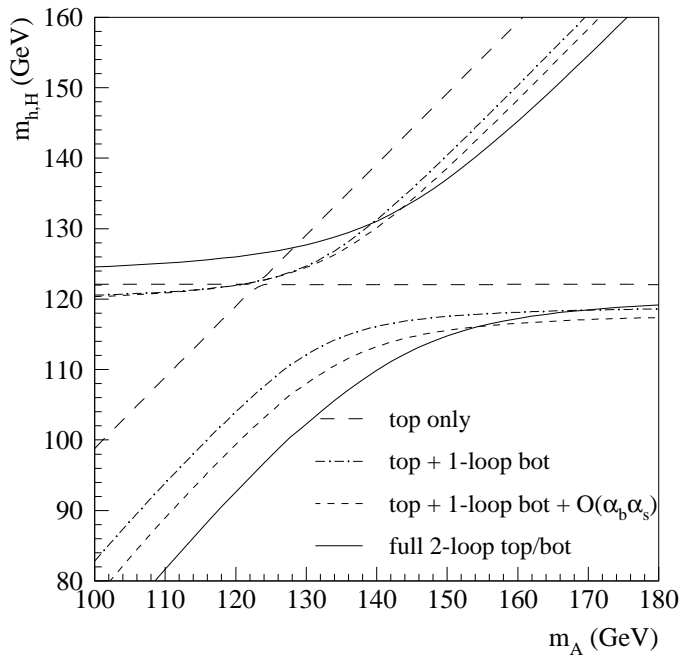,width=10cm,height=10cm}
\end{center}
\vspace{-0.5cm}
\caption{The masses $m_h$ and $\mH$ as a function of $\ma$, for $\tb =
 40,\, A_t = 1$ TeV, $A_b = 2$ TeV, $\mu = m_{Q,\tilde{t}} = m_U = m_D
 = \mgl = 1$ TeV. The meaning of the different curves is explained in
 the text.}
\label{mhhma}
\end{figure}
Finally, Fig.~\ref{mhhma} shows both CP--even Higgs masses, $\mh$ and
$\mH$, as functions of the CP--odd Higgs mass, in the region of
relatively small $\ma$ (100 GeV $< \ma <$ 180 GeV), for $\tb = 40$.
The other input parameters are chosen as $A_t = 1$ TeV, $A_b = 2$ TeV
$\mu = m_{Q,\tilde{t}} = m_U = m_D = \mgl = 1$ TeV. The meaning of the
various curves is the same as in Fig.~\ref{figmhtb}. Comparing the
short--dashed and solid curves we see that, for this choice of
parameters, the effect of the $\oatababq$ corrections is particularly
evident in the region where $\ma$ is small, and can account for
variations of several GeV (around 5 in this example) in both $\mh$ and
$\mH$.

In general, it appears from Figs.~\ref{figmhtb}--\ref{mhhma} that the
two--loop $\oabas$ and $\oatababq$ corrections are usually a small
fraction of the one--loop $\oab$ ones. We stress that this is a
desirable consequence of our renormalization prescription, which
allows to set apart the $\tan\beta$--enhanced threshold corrections,
resummed to all orders in the renormalized coupling $h_b$. If we were
to adopt for the bottom/sbottom sector the same renormalization
prescription that we use for the top/stop sector, the dependence on
$\tan\beta$ of the one--loop corrections would be smoother, but very
large corrections would appear at two loops, questioning the validity
of the perturbative expansion.

To conclude this section, we notice that our knowledge of the general
formulae for the corrections to the CP--even Higgs mass matrix in the
$\drbar$ scheme allows us to estimate the uncertainty connected with
the fact that we take the limit of
$\tan\beta\rightarrow\infty\,,\,\mb\rightarrow 0$ in the corresponding
OS results. In the numerical examples considered above we find that,
in the regions where the corrections are sizeable, the $\drbar$
results for the $\oabas$ part of $\left(\Delta{\cal M}^2_S\right)^{\rm
eff}$ vary by less than 20\% when the limit
$\tan\beta\rightarrow\infty\,,\,\mb\rightarrow 0$ is taken.  The
$\drbar$ results for the $\oatababq$ part of the corrections vary
instead by less than 10\%. We can assume that similar variations
occurr also in the corresponding OS results, which leads to shifts 
in $\mh$ typically smaller than 1 GeV.


\section{Conclusions and discussion}

In this paper we computed the $\oatababq$ corrections to the MSSM
neutral Higgs boson masses and to the minimization conditions of the
MSSM effective potential. Such corrections are relevant when the ratio
of the Higgs VEVs, $\tan\beta$, is large. Combined with the previously
computed $\oatas$ \cite{dsz}, $\oatq$ \cite{bdsz} and $\oabas$
\cite{bdsz2} corrections to the neutral Higgs masses, and with the
corresponding corrections to the minimization conditions of the
effective potential \cite{dstad}, these results provide us with a
complete computation of the leading two--loop corrections controlled
by the top and bottom Yukawa couplings. 

Using the formalism of the effective potential, we obtained complete
analytic expressions for the momentum--independent part of the
$\oatqatababq$ corrections, valid for arbitrary values of the MSSM
input parameters when the latter are expressed in the $\drbar$
renormalization scheme. We also discussed an OS renormalization
prescription for the parameters of the bottom/sbottom sector in the
physically relevant limit of large $\tan\beta$. Such prescription,
first introduced in Ref.~\cite{bdsz2} for the case of the $\oabas$
corrections, allows to separate the large threshold corrections
appearing in the relation between $h_b$ and the pole bottom mass from
the genuine two--loop effects. Finally, we discussed the numerical
impact of our results in a few representative examples, showing that,
for large values of $\tan\beta$, the $\oatababq$ corrections can
induce variations in the Higgs masses of the order of a few GeV. Since
our analytic expressions, both in the $\drbar$ and the OS schemes, are
too long to be useful if explicitly written on paper, we choose to
make them available in the form of a Fortran code.

Although the terms controlled by the top and bottom Yukawa couplings
undoubtedly account for the bulk of the two--loop corrections, several
pieces are still missing for a complete two--loop computation of the
MSSM Higgs masses. When $\tan\beta$ is large, the corrections
controlled by the tau Yukawa coupling $h_{\tau}$ might in principle be
non--negligible. In the approximation of neglecting the electroweak
gauge couplings, the only two--loop corrections involving the tau Yukawa 
coupling are those of $\oabatau$ and those of $\oatauq$, 
where $\atau = h_{\tau}^2/(4\pi)$. While the mixed
$\oabatau$ corrections would require a dedicated
computation~\footnote{We thank A.~Brignole for drawing our attention
on this point.}, explicit formulae for the $\oatauq$ corrections can
be obtained from the formulae of Refs.~\cite{bdsz,dstad} for the
purely $\oatq$ corrections, with the replacements $t \rightarrow
\tau\,,\; \tilde{b}_L \rightarrow \tilde{\nu}_{\tau}\,,\; N_c
\rightarrow 1$ and $H_1 \leftrightarrow H_2$ [the latter resulting
into $\tan\beta \leftrightarrow \cot\beta\,,\; v_1 \leftrightarrow
v_2\,,\; \left(\Delta{\cal M}^2_S\right)^{\rm eff}_{11}
\leftrightarrow \left(\Delta{\cal M}^2_S\right)^{\rm eff}_{22}$ and
$\sigu \leftrightarrow \sigd$]. If the input parameters are given in
the OS scheme, a suitable definition of $\delta A_{\tau}$ is required
in order to avoid introducing $\tan\beta$--enhanced terms in the
two--loop part of the result.  Anyway, we find that the $\oatauq$
corrections to the Higgs masses are in general very small compared
with those controlled by the bottom Yukawa coupling.  Besides the
hierarchy between $\mb$ and $m_{\tau}$, the suppression of the tau
corrections is motivated by the absence of color enhancements, and by
the fact that the only $\tan\beta$--enhanced threshold corrections to
the relation between $h_{\tau}$ and $m_{\tau}$ are those controlled by
the electroweak gauge couplings.

A full two--loop determination of the MSSM Higgs masses will require
going beyond the gaugeless limit and the effective potential
approximation, i.~e.~including both the corrections controlled by the
electroweak gauge couplings and the effect of the momentum--dependent
part of the Higgs self--energies. It can also be expected that, among
the three--loop corrections, at least those involving the top Yukawa
coupling affect the Higgs masses in a non--negligible way. In
Ref.~\cite{Martin2} the two--loop, zero--momentum electroweak
corrections have been computed numerically in a representative
scenario, and found to yield a shift in the lightest Higgs boson mass
$\mh$ of about 1 GeV with respect to the result obtained in the
gaugeless approximation. In Ref.~\cite{dhhsw}, the theoretical
uncertainty in the prediction for $\mh$ arising from the combined
effect of the missing two--loop corrections and the leading
three--loop corrections has been estimated to be around 3 GeV.

If the MSSM is a viable theory for physics at the weak scale, a light
Higgs boson will be discovered either at the Tevatron or at the LHC.
Subsequently, its properties will be determined with high precision at
a future linear collider: for example, the predicted experimental
accuracy in the determination of $\mh$ at TESLA is about 50 MeV
\cite{tesla}. It is thus clear that further effort will be required in
the coming years, in order to improve the accuracy of the theoretical
predictions up to the level required to compare with the experimental
results expected at the next generation of colliders.


\vspace{0.2cm}

\section*{Acknowledgments}

We would like to thank A.~Brignole, A.~Pilaftsis and F.~Zwirner for
useful comments and discussions, and T.~Hahn for help in producing the
Fortran routines.  This work was partially supported by the European
Community's Human Potential Programmes HPRN-CT-2000-00148 (Across the
Energy Frontier) and HPRN-CT-2000-00149 (Collider Physics).


\vspace{0.2cm}

\section*{Appendix}
\begin{appendletterA}

We present here the expressions for the functions $F_i^{t}\; (i =
1,2,3,4)\,,\; F_5\,,\; F_6\,,\; F^t\,,\; G^t$ and $F_A$, appearing in
Eqs.~(\ref{dms11})--(\ref{deltama}), in terms of derivatives of the
$\drbar$--renormalized $\Delta V$, computed at the minimum of $\Veff$:

\vspace{0.5cm}

\bea
F^t_1 & = & \DVtt + \DVtutu+\DVtdtd+ 2\,\DVttu+2\,\DVttd+2\,\DVtutd\nn\\
&&\nn\\
&+&  \frac{1}{4\,\mt^4} \,\left(
\DVcptpb + z_t\,\DVcptmptt + z_b\,\DVcptptb \right)\,,\\
&&\nn\\
&&\nn\\
F^t_2 & = & \DVtutu-\DVtdtd +\DVttu -\DVttd\nn\\
&&\nn\\ 
&-&\frac{4\,\Cdt^2}{\difft}\,\left(\DVcdtqt+\DVcdtqtu+\DVcdtqtd\right)
-\frac{z_t}{\Sdt^2\,\mt^2\,(\difft)} \,\DVcptmptt\,,\nn\\
&&\\
&&\nn\\
F^t_3 & = & \DVtutu+\DVtdtd - 2\,\DVtutd 
-\frac{2}{\difft}\left(\DVtu-\DVtd\right)\nn\\
&&\nn\\
&+&\frac{16\,\Cdt^2}{(\difft)^2}\,\left(\Cdt^2\,\DVcdtqcdtq+2\,\DVcdtq\right)
-\frac{8\,\Cdt^2}{\difft}\,\left(\DVcdtqtu-\DVcdtqtd\right)\nn\\
&&\nn\\
&+&\frac{4\,z_t}{\Sdt^4\,(\difft)^2}\left(
\DVcptmptt+\DVcpbptt + z_b\,\DVcpttptb \right)\,,\\
&&\nn\\
&&\nn\\
F^t_4 & = & \DVtub + \DVtubu + \DVtubd - \DVtdb - \DVtdbu - \DVtdbd \nn\\
&&\nn\\
&-& \frac{4\,\Cdt^2}{\difft} \,\left(\DVbuc2t + \DVbdc2t + \DVbc2t \right)
-\frac{z_t}{\Sdt^2\,\mb^2\,(\difft)}\, \DVcpbptt\,, \nn\\
&&\\\
&&\nn\\
F_5 & = & \DVtubu - \DVtubd - \DVtdbu + \DVtdbd \nn\\
&&\nn\\
&-& \frac{4\,\Cdt^2}{\difft} \,\left(\DVbuc2t - \DVbdc2t\right)
-\frac{4\,\Cdb^2}{\diffb} \,\left(\DVtuc2b - \DVtdc2b\right)\nn\\
&&\nn\\
&+& \frac{16 \, \Cdt^2 \Cdb^2}{(\difft)(\diffb)}\, \DVc2tc2b
- \frac{4\,z_t\,z_b}{\Sdt^2\,\Sdb^2\,(\difft)\,(\diffb)}\, \DVcpttptb\,,\nn\\
&&\\
&&\nn\\ 
F_6 & = & \DVtb + \DVtub + \DVtdb + \DVbut + \DVbdt  \nn\\
&&\nn\\
&+& \DVtubu + \DVtubd + \DVtdbu + \DVtdbd 
-\frac{1}{4\,\mt^2\,\mb^2}\,\DVcptpb\,,\\
&&\nn\\
&&\nn\\ 
&&\nn\\
\label{deff}
F^t & = & \frac{\partial \Delta V}{\partial \mstu}
- \frac{\partial \Delta V}{\partial \mstd}
-\frac{4\, \Cdt^2}{\difft} \,\frac{\partial \Delta V}{\partial \Cdt^2}\,,\\
&&\nn\\
&&\nn\\
\label{defg}
G^t & = & \frac{\partial \Delta V}{\partial m_t^2}
+ \frac{\partial \Delta V}{\partial \mstu}
+ \frac{\partial \Delta V}{\partial \mstd}\,,\\
&&\nn\\
&&\nn\\
&&\nn\\ 
\label{defFA}
F_A & = & \frac{1}{4\,\mt\,\mb}\,\DVcptpb
+ \frac{4\,(A_t\,A_b-\mu^2)^2\,\mt\,\mb\,z_t\,z_b}
{\Sdt^2\,\Sdb^2\,(\difft)^2\,(\diffb)^2}\,\DVcpttptb\nn\\
&&\nn\\
&+&\frac{\mt\,z_t}{\Sdt^2\,\mb\,(\difft)^2}\,
\left(A_t^2\,\DVcpbptt + \mu^2\,\cot^2\beta\,\DVcptmptt\right)\nn\\
&&\nn\\
&+&\frac{\mb\,z_b}{\Sdb^2\,\mt\,(\diffb)^2}\,
\left(A_b^2\,\DVcptptb + \mu^2\,\tan^2\beta\,\DVcpbmptb\right)\,.
\eea
\vspace{0.3cm}

\noindent In the above formulae, $z_q \equiv {\rm sign}(X_q)$,~\footnote{
Factors of $z_t$ were omitted in Eqs.~(28)--(30) and (C2) of
Ref.~\cite{dsz}. Notice also that in Eqs.~(\ref{deltama}) and
(\ref{defFA}) we employ a different definition of $F_A$ compared to
the one employed in Eqs.~(C1)--(C2) of Ref.~\cite{dsz}.}  where $X_q
\;\, (q = t,b)$ is the squark mixing parameter.  The functions
$F_i^{b}\,,\; F^b$ and $G^b$ can be obtained from their top
counterparts through the replacement $t \leftrightarrow b$.

\end{appendletterA}


\end{document}